\documentclass[sigplan,twocolumn]{acmart}
\usepackage{subfig}
\usepackage{booktabs, tabularx, makecell}
\usepackage[ruled,linesnumbered,noend]{algorithm2e}
\usepackage{pifont}
\usepackage{hyperref}

\acmSubmissionID{50}
\renewcommand\footnotetextcopyrightpermission[1]{} 

\AtBeginDocument{%
  }

\setcopyright{acmlicensed}
\copyrightyear{2018}
\acmYear{2018}
\acmDOI{XXXXXXX.XXXXXXX}
\acmConference[Conference acronym 'XX]{Make sure to enter the correct
  conference title from your rights confirmation emai}{June 03--05,
  2018}{Woodstock, NY}
\acmISBN{978-1-4503-XXXX-X/18/06}

\newcommand{\sysname}{{DeepCEE}\xspace}
\begin{document}


\title[\sysname: Fast Geo-Distributed Model Training in CEE Computing Environments]{\sysname: Efficient Cross-Region Model Distributed Training System under Heterogeneous GPUs and Networks}

\author{Jinquan Wang}
\authornote{Both authors contributed equally to this research.}
\email{derekjqwang@buaa.edu.cn}
\orcid{0000-0001-6690-8386}
\author{Xiaojian Liao}
\authornotemark[1]
\email{liaoxj@buaa.edu.cn}
\affiliation{%
  \institution{Beihang University}
  \city{Haidian}
  \state{Beijing}
  \country{China}
}

\author{Xuzhao Liu}
\affiliation{%
  \institution{Beihang University}
  \city{Haidian}
  \state{Beijing}
  \country{China}
}
\email{liuxuzhao@buaa.edu.cn}

\author{JiaShun Suo}
\affiliation{%
  \institution{Beihang University}
  \city{Haidian}
  \state{Beijing}
  \country{China}
}
\email{suojiashun@foxmail.com}

\author{Zhisheng Huo}
\authornote{Corresponed author.}
\affiliation{%
  \institution{Beihang University}
  \city{Haidian}
  \state{Beijing}
  \country{China}
}
\email{huozhisheng1122@126.com}

\author{Chenhao Zhang}
\affiliation{%
  \institution{Beihang University}
  \city{Haidian}
  \state{Beijing}
  \country{China}
}
\email{zch13021728086@buaa.edu.cn}

\author{Xiangrong Xu}
\affiliation{%
  \institution{Beihang University}
  \city{Haidian}
  \state{Beijing}
  \country{China}
}
\email{2037369569@qq.com}

\author{Runnan Shen}
\affiliation{%
  \institution{Beihang University}
  \city{Haidian}
  \state{Beijing}
  \country{China}
}
\email{shenrn@buaa.edu.cn}

\author{Xilong Xie}
\affiliation{%
  \institution{Beihang University}
  \city{Haidian}
  \state{Beijing}
  \country{China}
}
\email{xxl1399@buaa.edu.cn}

\author{Limin Xiao}
\authornotemark[2]
\affiliation{%
  \institution{Beihang University}
  \city{Haidian}
  \state{Beijing}
  \country{China}
}
\email{xiaolm@buaa.edu.cn}

\renewcommand{\shortauthors}{Wang et al.}

\begin{abstract}
Most existing training systems focus on  a single region. 
In contrast, we envision that cross-region training offers more flexible GPU resource allocation and yields significant potential.
However, the hierarchical cluster topology and unstable networks in the cloud-edge-end (CEE) environment, a typical cross-region scenario, pose substantial challenges to building an efficient and autonomous model training system.
We propose \sysname, a geo-distributed model training system tailored for heterogeneous GPUs and networks in CEE environments. 
\sysname adopts a communication-centric design philosophy to tackle challenges arising from slow and unstable inter-region networks. 
It begins with a heterogeneous device profiler that identifies and groups devices based on both network and compute characteristics. 
Leveraging device groups, \sysname implements compact, zero-bubble pipeline parallelism, automatically deriving optimal parallel strategies. 
To further adapt to runtime variability, \sysname integrates a dynamic environment adapter that reacts to network fluctuations.
Extensive evaluations demonstrate that \sysname achieves 1.3-2.8$\times$ higher training throughput compared to widely used and SOTA training systems.

\end{abstract}

\received{20 February 2007}
\received[revised]{12 March 2009}
\received[accepted]{5 June 2009}

\maketitle

\section{Introduction}
\label{introduction}
The advancement of AI technology \cite{heDeepResidualLearning2016, vaswaniAttentionAllYou2023, daiTransformerXLAttentiveLanguage2019, kitaevReformerEfficientTransformer2020, munkhdalaiLeaveNoContext2024, redmonYouOnlyLook2016, huangDenselyConnectedConvolutional2017} is driving progress across various industries \cite{douPotentialAIService2024, meloniCloudbasedIoTSolution2018, watanabeDynamicMap20Traffic2020}, with model training being one of the most critical components. 
Traditionally, model training is conducted within a single cloud data center. 
However, we envision that cross-region distributed training \cite{liuCrossdomainRandomPretraining2025, zhongStreamRLScalableHeterogeneous2025a} will become a key direction for reducing both the barriers and costs associated with model training \cite{bhardwajEkyaContinuousLearning2022, miaoUnifiedReplayBasedContinuous2024, yeAsteroidResourceEfficientHybrid2024, xuSoCFlowEfficientScalable2024}, for the following reasons.

\begin{figure}[t]
    \centering

    \subfloat[The workload of the cloud datacenter and the edge/end within 24 hours in real world \cite{wengMLaaSWildWorkload2022, liELASTICEdgeWorkload2023}.]{\includegraphics[width=4cm,height=2.5cm]{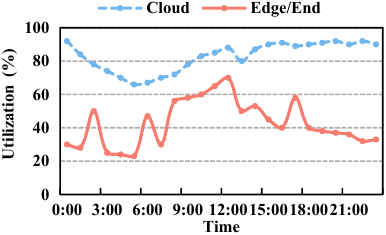}\label{fig:ecosystem:workload}}
    \hfill
    \subfloat[The CEE environment, a typical cross-region scenario.]{\includegraphics[width=4cm]{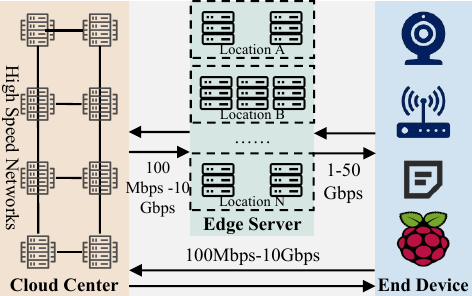}\label{fig:ecosystem:ecosystem}}

    \vspace{-0.25cm}
    \caption{The CEE environment and its workload status.}
    \vspace{-0.5cm}
    \Description{The Cloud-Edge-End computing3 ecosystem.}
    \label{fig:ecosystem}
\end{figure}

First, as model sizes continue to grow, the demand for GPUs is also increasing \cite{zhou2024TrainingServingSystem, jiang2024MegaScaleScalingLarge, teamGemini15Unlocking2024}.
Training tasks cannot be executed immediately  \cite{tianBreakingMemoryWall2024, tamFedHybridBreakingMemory2024, wangSwapNetEfficientSwapping2024} when GPUs in a cloud data center are fully utilized.
We observe this phenomenon in real-world production environments (Figure~\ref{fig:ecosystem:workload}): the daily load of a cloud datacenter can reach as high as 78\%, with more than 14 hours a day operating above 90\%.

Second, physical constraints such as land availability, pow-er consumption, and energy supply are making it increasingly difficult to expand a single cloud data center \cite{MultiDatacenterTrainingOpenAIs2024, teamGemini15Unlocking2024}.
As a result, the mainstream approach has shifted toward deploying multiple cloud and edge data centers, with some end devices equipped with GPU resources
(e.g., Nvidia P4 \cite{nvidiaTESLAP4GPU2025}, RTX 3000 series \cite{NVIDIAGeForceRTX}, RTX 4000 series \cite{GeForceRTX4090}, and Cambricon MLU 200 \cite{zhangCambriconXAcceleratorSparse2016} etc.)
However, we observe that edge-side computing capabilities remain significantly underutilized, with average GPU utilization rates as low as 28\% \cite{yeAsteroidResourceEfficientHybrid2024, kwonTinyTrainResourceAwareTaskAdaptive2024}.
This presents a promising opportunity to harness idle GPU resources at the edge for model training.

Third, with the continuous advancement of networking technologies, the bandwidth between different regions has significantly increased, making efficient cross-region model training increasingly feasible.
As shown in Figure~\ref{fig:ecosystem:ecosystem}, the cloud center, edge server and end device are connected by high-speed WANs whose bandwidth is between 100Mbps to 10Gbps~\cite{cheikhMultilayeredEnergyEfficiency2022, chilamkurthyLowpowerWideareaNetworks2022, liReasoningNetworkTraffic2024, LocalAreaNetwork2025}.
We refer to this environment formed by high-speed networks 
\cite{hanDynamicMultifacetedSpatiotemporal2021, tangEasySTSimpleFramework2024, zhangTrafficgptViewingProcessing2024} and heterogeneous computing resources as the CEE environment.

\begin{figure}[t]
    \centering

    \subfloat[Throughput]{\includegraphics[width=2.7cm,height=2.7cm]{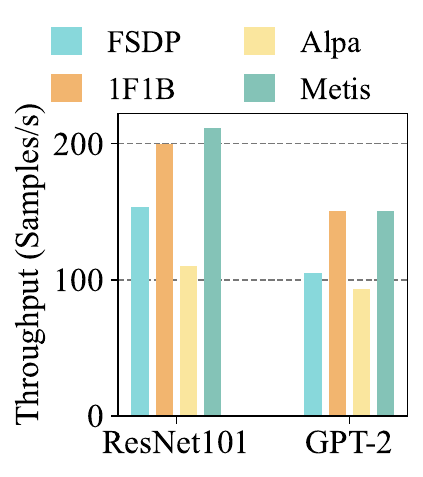}\label{fig:parallelism-methods:pp}}
    \hfill
    \subfloat[Utilization]{\includegraphics[width=2.7cm,height=2.7cm]{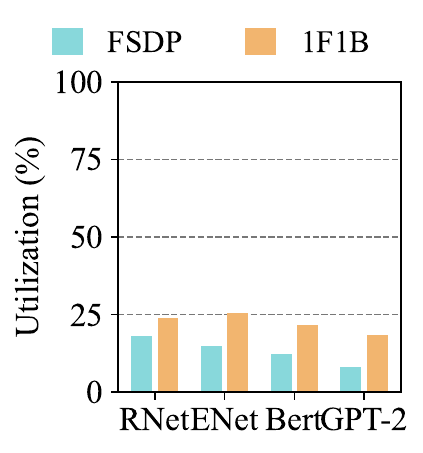}\label{fig:parallelism-methods:util}}
    \hfill
    \subfloat[Time Proportion]{\includegraphics[width=2.7cm,height=2.7cm]{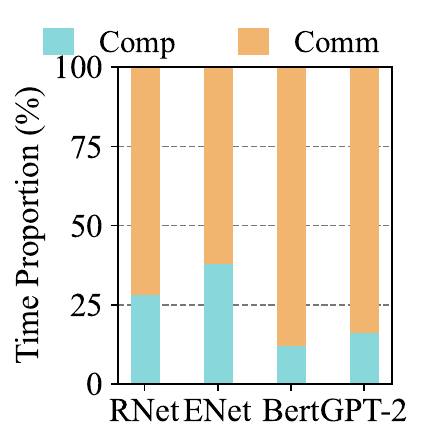}\label{fig:parallelism-methods:dp}}

    \vspace{-0.25cm}
    \caption{Performance comparison of different methods in a real CEE environment.}
    \vspace{-0.5cm}
    \Description{Performance comparison of different methods in the real Cloud-Edge-End computing ecosystem.}
    \label{fig:parallelism-methods}
\end{figure}

Despite its strong potential, cross-region training still encounters significant technical challenges related to automation and efficiency.
We start by directly extending existing training frameworks to the CEE environment.
As shown in Figure~\ref{fig:parallelism-methods:pp}, we first apply commonly used training frameworks such as PyTorch FSDP \cite{liPyTorchDistributedExperiences2020, zhaoPyTorchFSDPExperiences2023} and DeepSpeed 1F1B \cite{MicrosoftDeepSpeedDeepSpeed, smithUsingDeepSpeedMegatron2022} to cross-region training. 
However, we observe that the training throughput was significantly lower than that of single-region training, with excessive pipeline bubbles leading to GPU utilization falling below 25.6\% (Figure~\ref{fig:parallelism-methods:util}).

We further deploy state-of-the-art training frameworks (e.g., Alpa \cite{zhengAlpaAutomatingInter2022}, Metis \cite{umMetisFastAutomatic2024}) that can automatically generate training strategies suitable for the heterogeneous devices in CEE environments.
However, we observe that these frameworks still fail to fully utilize available GPU resources, with training throughput significant declining.
Worse still, the execution plan generation process in some systems (e.g., Asteroid \cite{yeAsteroidResourceEfficientHybrid2024}) introduces substantial time overhead, ultimately prolonging the overall training duration.
These findings suggest that existing training frameworks fall short in effectively leveraging the aggregated computing capabilities across the CEE environment.

Through in-depth analysis (\S\ref{challenges}), we identify the root cause as the excessive communication overhead (Figure~\ref{fig:parallelism-methods:dp}), which could not be effectively overlapped with computation.
This issue stems from two unique characteristics of the CEE environment.
First, the environment features a hierarchical cluster topology composed of highly heterogeneous networks and GPUs.
As shown in Figure~\ref{fig:ecosystem:ecosystem}, in addition to high-speed RDMA networks within cloud data centers, the CEE environment includes high-speed LANs connecting edge servers and relatively slow WANs bridging cloud centers, edge clusters, and end devices.
The average bandwidth of these WANs varies significantly across segments, resulting in a complex network topology that is largely overlooked by existing training frameworks.
Moreover, the heterogeneity of GPU resources in the CEE environment is even more pronounced.
Table~\ref{tab:gpu-device} and Table~\ref{tab:gpu-capacity} summarize the GPU configurations across the cross-region setup, encompassing at least 7 different GPU models.
These GPUs differ substantially in terms of compute capabilities, memory capacity and bandwidth, making it even more challenging to generate optimal parallelization strategies automatically.

\begin{table}[t]
    \centering
    \footnotesize
    \caption{Heterogeneous GPU deployment in Alibaba Cloud and common GPU distribution patterns within edge servers and end devices.}
    \vspace{-0.25cm}
    \begin{tabularx}{\linewidth}{ccccccc}
        \toprule
        \textbf{Region} & \textbf{P100} & \textbf{A10} & \textbf{V100} & \textbf{P4} & \textbf{RTX4000} & \textbf{RTX3000} \\
        \midrule
        Qingdao Cloud & \ding{55} & \ding{55} & \ding{52} & \ding{52} & \ding{55} & \ding{55} \\
        Beijing Cloud & \ding{52} & \ding{52} & \ding{52} & \ding{52} & \ding{55} & \ding{55} \\
        Hohhot Cloud & \ding{52} & \ding{55} & \ding{55} & \ding{55} & \ding{55} & \ding{55} \\
        Edge Server & \ding{55} & \ding{55} & \ding{55} & \ding{52} & \ding{52} & \ding{55} \\
        End Device & \ding{55} & \ding{55} & \ding{55} & \ding{55} & \ding{52} & \ding{52} \\
        \bottomrule
    \end{tabularx}
    \label{tab:gpu-device}
\end{table}

\begin{table}[]
    \centering
    \footnotesize
    \caption{Specifications of heterogeneous GPUs capabilities, including computing, memory, and power consumption.}
    \vspace{-0.25cm}
    \begin{tabularx}{\linewidth}{X<{\centering}ccccc}
        \toprule
        \textbf{GPU} & \textbf{\makecell*[c]{FP16\\(TFLOPs)}} & \textbf{\makecell*[c]{FP32\\(TFLOPs)}} & \textbf{\makecell*[c]{Memory\\(GB)}} & \textbf{\makecell*[c]{Bandwidth\\(GB/s)}} & \textbf{\makecell*[c]{Power\\(W)}}  \\
        \midrule
        \textbf{P100} & 18.7 & 9.3 & 16 & 732 & 250 \\
        \textbf{A10} & 62.4 & 31.2 & 24 & 600 & 150 \\
        \textbf{A100} & 312 & 156 & 40 & 1555 & 400 \\
        \textbf{P4} & 5.5 & / & 8 & 192 & 75 \\
        \textbf{RTX4090} & 82.58 & 41.2 & 24 & 1018 & 425 \\
        \textbf{RTX3090} & 71.16 & 35.58 & 24 & 936.2 & 350 \\
        \textbf{RTX3080} & 59.54 & 29.77 & 10 & 760 & 320 \\
        \bottomrule
    \end{tabularx}
    \vspace{-0.5cm}
    \label{tab:gpu-capacity}
\end{table}

Second, in the CEE environment, training tasks must contend with other workloads (e.g., bursty tasks and routine data transfers \cite{zhengPersonalizedElasticEmbedding2024, xiaAerorecEfficientOndevice2024, yangDiscreteFederatedMultibehavior2024, zhangOptimalContainerUpdate2024}) for cross-region network resources, resulting in bandwidth fluctuations and network instability during training.
Our experimental results using real-world workloads demonstrate that merely launching a large data transfer task during training can cause resource contention, reducing throughput by over 56\% (details in Figure~\ref{fig:adapter-exp:throughput}).
Current training frameworks fail to adapt to dynamic network bandwidth conditions, thus preventing efficient training in the CEE environment.

We present \sysname (\S\ref{span-train}), a geo-distributed model training system that can automatically generate parallel strategies and adapt to network fluctuations in the CEE environment.
The core idea of \sysname is network-centric automatic parallel strategy generation and adjustment. 
While most existing work focuses on the impact of compute heterogeneity on parallel strategies, in CEE environments, network heterogeneity becomes the primary factor affecting training efficiency. 
This makes it difficult for existing approaches to fully utilize GPU compute resources across regions. 
\sysname introduces a network-centric approach, prioritizing the mitigation of pipeline bubbles caused by cross-region networks in device grouping, automatic parallel strategy generation, and the adjustment of pipeline stages during execution.
Specifically, \sysname includes the following three key techniques.

First, \sysname introduces the Heterogeneous Devices Profiler, which plans a two-level device grouping before training begins. 
The first-level groups GPUs based on network performance, eliminating the impact of cross-region network hierarchies on parallel training strategies. 
The second level further groups GPUs within each network-isolated group based on compute capability, clustering GPUs with similar computation and memory access performance.

Second, \sysname proposes the Parallel Strategy Planner, which adopts different parallel strategies for each level of the two-level device group. 
At the first level, considering the low bandwidth and variability of cross-region networks, \sysname introduces compact zero-bubble pipeline parallelism, where weight computations during back propagation are decoupled and inserted into potential pipeline bubbles in a compact fashion. 
At the second level, \sysname automatically generates the optimal combination of pipeline, data and tensor parallelism by jointly considering differences in compute and memory access capabilities across GPUs.

\begin{figure*}
    \centering
    
    \subfloat[Issues of hierarchical heterogeneous network when applying DP in CEE environment.]{\includegraphics[width=5.5cm,height=3.5cm]{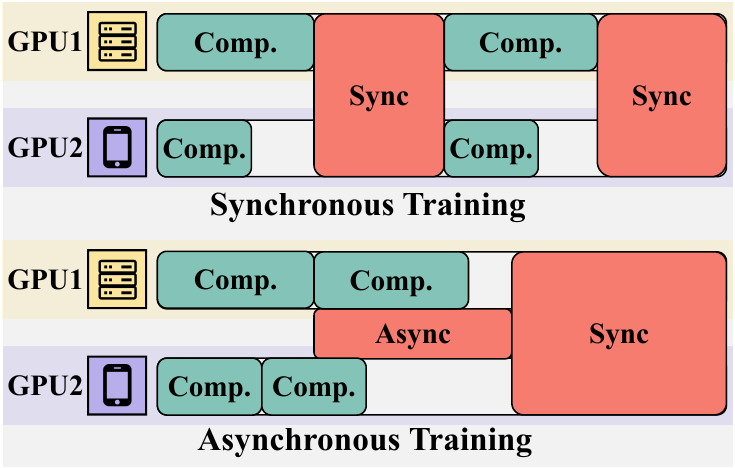}\label{fig:challenge:device}}
    \hfill
    \subfloat[Issues of hierarchical heterogeneous network when applying PP in CEE environment.]{\includegraphics[width=5.5cm,height=3.5cm]{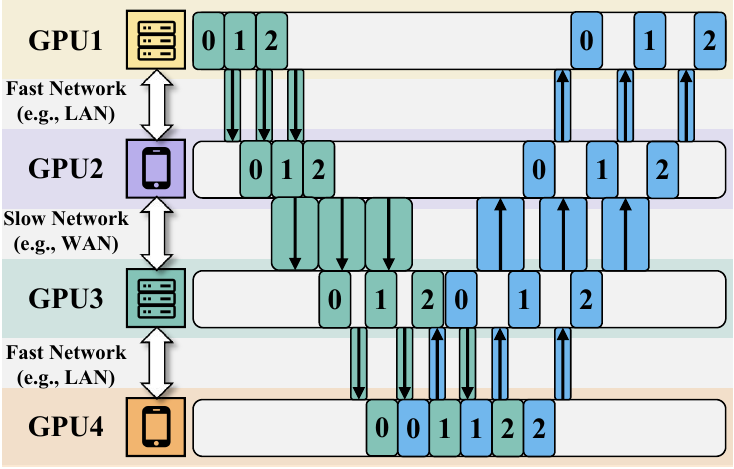}\label{fig:challenge:network}}
    \hfill
    \subfloat[Issues of network fluctuation when applying PP in CEE environment.]{\includegraphics[width=5.5cm,height=3.5cm]{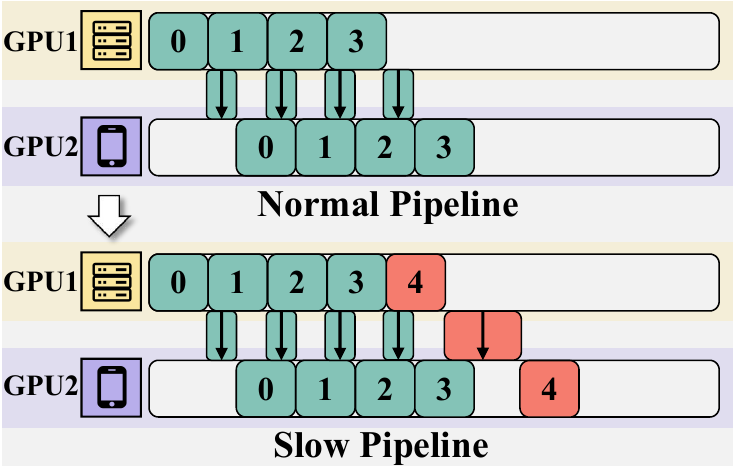}\label{fig:challenge:fluctuation}}

    \vspace{-0.3cm}
    \caption{In the CEE environment, current distributed training strategies face two challenges. The hierarchical cluster topology fundamentally constrains training throughput in the CEE environment, while network fluctuations frequently triggers abrupt performance degradation.}
    \vspace{-0.3cm}
    \Description{}
    \label{fig:challenge}
\end{figure*}

Third, \sysname introduces the Dynamic Environment Adapter to cope with network fluctuations.
Unlike existing methods that rely on checkpoints and pipeline restarts, \sysname monitors network variability and adapts the system through micro-batch-level adjustments, enabling continuous and efficient training without interruptions.

We implement \sysname based on PyTorch with around 15K lines of code.
We evaluate \sysname against a wide range of state-of-the-art training systems including PyTorch DDP \cite{liPyTorchDistributedExperiences2020}, PyTorch FSDP \cite{zhaoPyTorchFSDPExperiences2023}, DeepSpeed ZeRO2 \cite{rajbhandariZeROMemoryOptimizations2020a}, DeepSpeed Pipeline \cite{narayananPipeDreamGeneralizedPipeline2019}, Gpipe \cite{huangGPipeEfficientTraining2019}, Alpa \cite{zhengAlpaAutomatingInter2022}, Metis \cite{umMetisFastAutomatic2024}, Hetpipe \cite{umMetisFastAutomatic2024}, and Asteroid \cite{yeAsteroidResourceEfficientHybrid2024}.
Extensive results (\S\ref{evaluation}) in a simulated CEE environment show that the \sysname outperforms existing distributed training systems by 1.3-2.8$\times$ in terms of training throughput.
Furthermore, under dynamic network changes, \sysname improves training throughput by 1.5-1.7$\times$ compared to existing systems.

In summary, we make the following contribution:

\noindent \setlength{\hangindent}{1em} \textbullet \ We analyze the potential benefits and challenges of extending model training to CEE computing environments.

\noindent \setlength{\hangindent}{1em} \textbullet \ We propose \sysname, a geo-distributed training system designed specifically for CEE computing environments, with a network-centric design philosophy at its core.

\noindent \setlength{\hangindent}{1em} \textbullet \ We implement \sysname based on PyTorch and conduct extensive evaluations to verify its effectiveness.



\section{Background and Motivation}
\label{challenges}

As shown earlier in \S\ref{introduction} and Figure~\ref{fig:parallelism-methods}, existing model training frameworks fail to efficiently utilize computing resources in the CEE environment.
In this section, we provide a detailed analysis of the underlying inefficiencies through concrete examples.
Our analysis focuses on two primary challenges arising from cross-region deployment: hierarchical cluster topology (\S\ref{hierarchical-topologies}) and network fluctuations (\S\ref{network-fluctuations}).

\subsection{Challenges 1: Hierarchical Cluster Topology}
\label{hierarchical-topologies}


As noted in \S\ref{introduction}, CEE environments involve a mix of network types, including relatively slow WANs.  
Given this, network efficiency becomes an important consideration when designing model training systems.
Among the commonly used strategies, Data Parallelism (DP), Tensor Parallelism (TP), and Pipeline Parallelism (PP), TP typically incurs the highest communication overhead~\cite{lepikhinGShardScalingGiant2020, shoeybiMegatronLMTrainingMultiBillion2020, bianMaximizingParallelismDistributed2021a, parkHetPipeEnablingLarge2020, narayananPipeDreamGeneralizedPipeline2019, fanDAPPLEPipelinedData2021, liFold3DRethinkingParallelizing2023, linNnScalerConstraintguidedParallelization2024}, and may be less favourable in bandwidth-constrained scenarios.
Therefore, we narrow our focus to DP and PP as potentially more suitable options for CEE environments.
To explore their efficiency in CEE environments, we begin by analyzing traditional DP (Figure~\ref{fig:challenge:device}) and 1F1B PP (Figure~\ref{fig:challenge:network}).


Figure~\ref{fig:challenge:device} illustrates common issues in DP model training across heterogeneous network in the CEE environment.
GPU1 and GPU2 represent two GPU devices with different computing capabilities, located in cloud and edge clusters, respectively.
Under synchronous training (e.g., DDP \cite{liPyTorchDistributedExperiences2020}, FSDP \cite{zhaoPyTorchFSDPExperiences2023}, ZeRO \cite{rajbhandariZeROMemoryOptimizations2020a}), 
although these two GPUs have different computational speeds, both of them can complete the computation process quickly.
However, after each training epoch, they must rely on slow cross-region network for synchronization.
This causes communication time to occupy a significant portion of the total training time, limiting training throughput.
Under asynchronous training (e.g., Megatron \cite{smithUsingDeepSpeedMegatron2022}, MindSpore \cite{MindSporeOfficialSite}), GPU1 can proceed to the next epoch without waiting for weight synchronization.
However, to maintain model convergence, asynchronous updates are typically limited to being at most one epoch ahead.
When the cross-region communication time is too large compared to the computing time, even the next epoch of training cannot overlap with communication, resulting in prolonged waiting periods that reduce efficiency.

\begin{figure*}
    \centering
    \includegraphics[width=16.5cm]{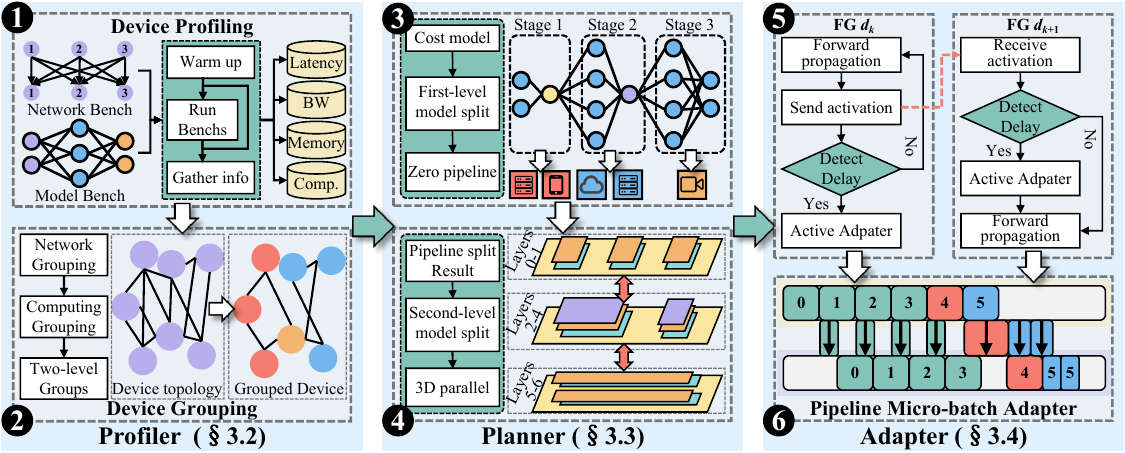}
    \vspace{-0.3cm}
    \caption{
        The main working components and workflow of \sysname.
        \sysname includes the pre-run performance evaluation component \textit{Heterogeneous Devices Profiler}, the pre-run parallel planning component \textit{Parallel Strategy Planner}, and the runtime environment adaptation component \textit{Dynamic Environment Adapter}.
    }
    \vspace{-0.4cm}
    \Description{
        The main working components and workflow of \sysname.
        \sysname includes the pre-run performance evaluation component \textit{Heterogeneous Devices Profiler}, the pre-run parallel plan planning component \textit{Parallel Strategy Planner}, and the runtime environment adaptation component \textit{Dynamic Environment Adapter}.
    }
    \label{fig:overview}
\end{figure*}

As shown in Figure~\ref{fig:challenge:network}, when implementing pipeline parallelism (e.g., HetPipe \cite{parkHetPipeEnablingLarge2020}, Asteroid \cite{yeAsteroidResourceEfficientHybrid2024}, AutoSF \cite{yangAutoSFAdaptiveDistributed2024}, FTPipeHD \cite{chenFTPipeHDFaultTolerantPipelineParallel2024}, Metis \cite{umMetisFastAutomatic2024}) across regions, the slowest network link becomes the critical bottleneck that severely limits training throughput.
In this scenario, GPU1-GPU2 and GPU3-GPU4 pairs enjoy high-speed connections (e.g., RDMA/LAN), while GPU2-GPU3 suffers from low-speed connections (e.g., WAN).
When GPU3 completes micro-batch \#0’s computing and transfers activations to GPU4, GPU2 simultaneously finishes micro-batch \#1, but cannot promptly transfer its activations to GPU3 due to the network bottleneck.
This forces GPU3 into an idle state, resulting in unavoidable pipeline bubbles.
The same issue arises during backward propagation,
negatively impacting the utilization of devices connected by high-speed networks (GPU1 / GPU2) and ultimately degrading the pipeline's efficiency.
The training performance is constrained by the slowest cross-region connection, exemplifying the "weakest link" problem.

Worse still, the computing heterogeneity (Tables~\ref{tab:gpu-device} and \ref{tab:gpu-capacity}) inherent in CEE environments further exacerbates the inefficiencies of traditional DP and PP caused by slow and heterogeneous networks.

Based on the above analysis, we conclude that both traditional DP and PP suffer from significant efficiency issues in cross-region CEE environments. 
However, we also observe that, compared to DP, the GPU bubbles in PP are more amenable to mitigation through pipeline optimization.

\subsection{Challenges 2: Network Fluctuations}
\label{network-fluctuations}

Network fluctuations are common in the CEE environment.
For example, multiple edge servers and end devices share uplinks to cloud datacenter.
But, bursty traffic patterns, including periodic data reporting and inference result transmission, can suddlenly dominate available bandwidth, resulting severe network fluctuations for training task.
Prior work \cite{liPyTorchDistributedExperiences2020, zhaoPyTorchFSDPExperiences2023, huangGPipeEfficientTraining2019, MicrosoftDeepSpeedDeepSpeed} is primarily designed for data center networks, under the assumption of relatively stable bandwidth.
Consequently, these approaches are ill-suited to handle the dynamic and unpredictable network conditions typical of the CEE environment.


Considering that pipeline parallelism imposes the least demand on network bandwidth among training parallelism techniques, 
we use a pipeline training example to demonstrate how network fluctuation degrades cross-region training efficiency.
In Figure~\ref{fig:challenge:fluctuation}, GPU1 and GPU2 each handle one pipeline stage.
Under stable network conditions ("Normal Pipeline" in Figure~\ref{fig:challenge:fluctuation}), computation fully overlaps with communication, creating a tightly packed pipeline.
However, when network bandwidth between GPU1 and GPU2 fluctuates and drops, communication delays can no longer be hidden by computation.
When GPU1 completes micro-batch \#4 (red block), the activation transfer to GPU2 fails to finish during GPU3’s computing of micro-batch \#3 ("Slow Pipeline" in Figure~\ref{fig:challenge:fluctuation}).
This forces GPU2 and downstream stages into idle states, significantly reducing hardware utilization.
Current pipeline systems lack adaptive mechanisms for such scenarios, remaining passive until network conditions recover, which often leads to prolonged throughput degradation during network fluctuations.

Existing approaches (e.g., AutoSF \cite{yangAutoSFAdaptiveDistributed2024}) employ checkpoint-based mechanisms to handle network fluctuations in CEE environments, require restarting the entire pipeline, which incurs prohibitive costs and significant reaction latency.

\section{\sysname: Design and Implementation}
\label{span-train}

We present \sysname, a cross-region model training system designed to address the unique challenges of the CEE environment, as discussed in \S\ref{challenges}.
This section first provides an overview (\S~\ref{overview}) and highlights the core idea behind \sysname, then delves into the key techniques, namely \textit{Heterogeneous Devices Profiler} (\S~\ref{heterogeneous-devices-profiler}), \textit{Parallel Strategy Planner} (\S~\ref{parallel-strategy-planner}), and \textit{Dynamic Environment Adapter} (\S~\ref{dynamic-environment-adapter}).

\subsection{Key Idea and Overview}
\label{overview}

Departing from the compute-centric approach of existing model training systems, \sysname generates and adapts parallelization strategies driven by network performance.
In the CEE environment, where the cluster topology exhibits a hierarchical characteristic in terms of computing device and network, in addition, network performance also varies significantly across different tiers and exhibits noticeable fluctuations (as shown in \S\ref{challenges}). 
Existing systems often mistakenly group cross-region devices into a single device group, resulting in suboptimal parallelization strategies and increased pipeline bubbles during network fluctuations. 
By treating network performance evaluation as a first-class citizen, \sysname effectively addresses the network challenges inherent in cross-region training, with employing computing device performance evaluation as a second-class citizen.

Figure~\ref{fig:overview} presents the overview and workflow of \sysname.
\sysname first employs the Profiler (\S\ref{heterogeneous-devices-profiler}) to find optimal device groups.
It then uses the Planner (\S\ref{parallel-strategy-planner}) to automatically generate an asymmetric hybrid parallel strategy.
Finally, during periods of network fluctuation, it introduces Adapter (\S\ref{dynamic-environment-adapter}) to mitigate pipeline bubbles, thereby ensuring stable training performance.
Specifically, the core components of \sysname are as follows:

\noindent \setlength{\hangindent}{1em} \textit{\textbullet \ Heterogeneous Devices Profiler (\S~\ref{heterogeneous-devices-profiler}).}
It performs lightweight, end-to-end performance profiling of heterogeneous computing devices and networks.
Based on the profiling results (e.g., network latency and bandwidth), it runs a device grouping algorithm to construct a two-level hierarchy: a first-level network device group and a second-level computing device group.
Notably, the first-level network device group is carefully designed to address the challenges of cross-region network environments.

\noindent \setlength{\hangindent}{1em} \textit{\textbullet \ Parallel Strategy Planner (\S~\ref{parallel-strategy-planner}).}
It introduces an asymmetric hybrid parallel mechanism based on the two-level device group.
Among various training parallelism strategies, pipeline parallelism offers the lowest network overhead, making it the preferred choice for communication across the first-level device groups in \sysname.
Given the slow cross-region networks, conventional 1F1B (one-forward-one-backward) pipeline execution introduces significant pipeline bubbles, resulting in suboptimal compute utilization.
To alleviate this inefficiency, \sysname adopts zero pipeline \cite{qiZeroBubblePipeline2023}, which separates weight computations during backpropagation and fills idle execution slots, thereby improving compute resource utilization under slow network conditions.
In the second-level device group, \sysname autonomously derives a 3D parallelism strategy guided by the principle of load balancing across heterogeneous compute resources. 
This involves sequentially searching for the optimal combination of pipeline, data, and tensor parallelism.

\noindent \setlength{\hangindent}{1em} \textit{\textbullet \ Dynamic Environment Adapter (\S \ref{dynamic-environment-adapter}).}
It continuously monitors fluctuations in network performance.
Upon detecting significant network-induced delay, e.g., when a later pipeline stage is stalled while waiting for activations, it activates a dynamic micro-batch adjustment mechanism. 
This mechanism reduces the micro-batch size of the preceding pipeline stage, enabling the subsequent stage to commence computation earlier and thereby enhancing pipeline utilization.

\subsection{Heterogeneous Devices Profiler}
\label{heterogeneous-devices-profiler}

As discussed in \S~\ref{hierarchical-topologies}, CEE exhibits distinct hierarchical cluster topology where significant variations exist in both network and device across regions.
Our observations reveal that inter-region network performance govern cross-region training performance, often becoming the critical bottleneck.
To address this issue, \textit{Profiler} automatically generate two-level device groups using a performance evaluation mechanism and a two-level device grouping mechanism.


\subsubsection{The Performance Evaluation Mechanism}

\textit{Profiler} initiates network performance evaluation by assessing bandwidth and latency between all device pairs.
Specifically, \textit{Profiler} generates two distinct test packets: (1) a large payload $\mathbb{P}_b$ sized according to each device's GPU memory capacity for bandwidth measurement, and (2) a minimal payload $\mathbb{P}_l$ containing just five \textit{int32} values for latency profiling.
Then, \textit{Profiler} performs multi-round network warmup using payload $\mathbb{P}_b$.
Upon completion, \textit{Profiler} executes CCL primitives (e.g., AllReduce or broadcast) to repeatedly transfer both $\mathbb{P}_b$ and $\mathbb{P}_l$, recording their average transfer times as $alpha$ and $\beta$ respectively.
These metrics are combined through the formula $p_t = \alpha + \beta / m$ to quantify the communication capability between any device pair, where $m$ represents $\mathbb{P}_b$'s size, establishing a unified performance metric that accounts for both bandwidth and latency characteristics.



Next, \textit{Profiler} executes lightweight model benchmarks comprising diverse neural network architectures (e.g., CNN, Transformer, MLP) and mixed-precision matrix operations to assess heterogeneous devices' computing capabilities.
Benchmarks are run repeatedly, recorded average execution time $t_i$ for each task.
Subsequently, \textit{Profiler} calculates each device's computing capacity using the formula $p_c = \sum_i (\frac{w_i}{t_i})$, where $w$ represents task-specific weights, quantifying its overall processing capability for parallel workload assignment.



\subsubsection{The Two-level Device Grouping Mechanism}

\begin{figure}[t]
    \centering
    \includegraphics[width=7cm]{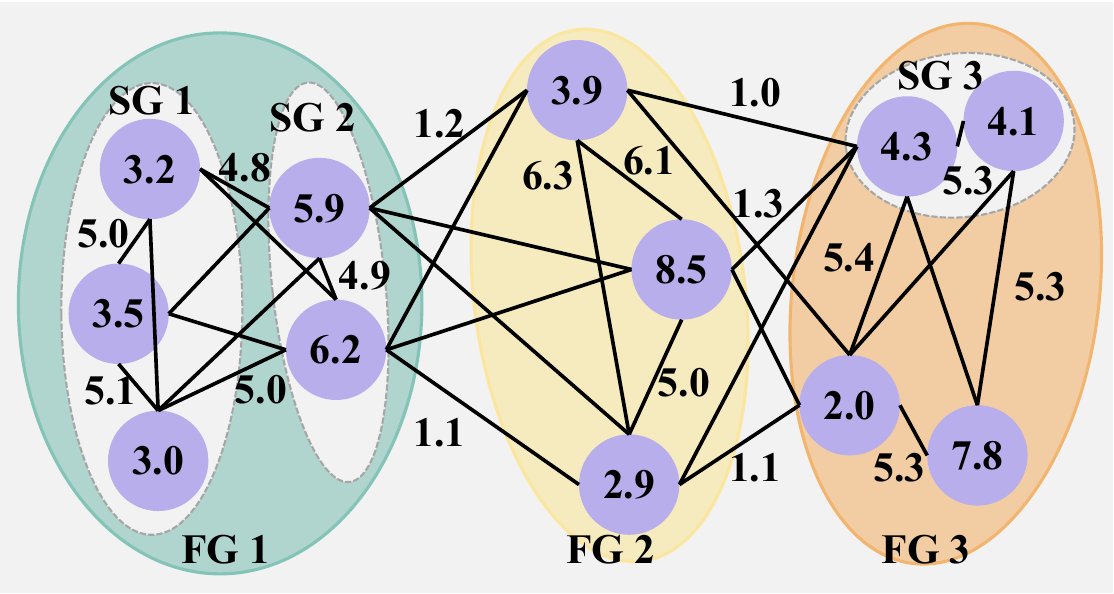}
    \caption{An example of heterogeneous device and network grouping in the CEE environment. 
    (FG: first-level network device group; SG: second-level computing device group)}
    \vspace{-0.6cm}
    \Description{Simple CEE heterogeneous device grouping. Circles represent computing devices, while numbers indicate their computing capabilities. The lines between the circles represent the network between the devices, while the numbers next to them show network capabilities. (FG: first-level network device group; SG: second-level computing device group)}
    \label{fig:device-group-example}
\end{figure}

Departing from prior cross-region approaches that group devices at node-level granularity, \textit{Profiler} implements a device-level fine-grained GPU grouping mechanism akin to data center practices.
\textit{Profiler} performs hierarchical clustering based on network capabilities to establish first-level network device groups, adapting CEE's hierarchical cluster topology.
Initially, each device is treated as an independent group, with inter-device communication metrics stored in a max-heap $v_n$.
The first-level network device group algorithm then iteratively (1) extracts the top element from $v_n$ to identify candidate groups, (2) merges groups if the difference between their network capability values is below a predefined threshold, and (3) updates $v_n$ by incorporating new group and computing their average network metrics, gradually constructing CEE's cross-region topology through network-aware aggregation.
Notably, \textit{Profiler} retains only the top-level groupings after clustering, discarding intermediate subgroups formed during the algorithm process.
As shown in Figure~\ref{fig:device-group-example}, this approach partitioned 12 devices into 3 first-level network device group, each exhibiting internally high bandwidth while clearly delineating cross-region boundaries.

Within each first-level network device group, \textit{Profiler} executes a second-level computing device grouping algorithm, similar to the first-level network device group algorithm but optimized for computing capabilities, to further partition devices into compute-homogeneous second-level computing groups.
Figure~\ref{fig:device-group-example} illustrates three representative outcomes: (1) FG1 splits into three subgroups (SG1-3), (2) FG2 maintains all devices as individual compute units, and (3) FG3 forms hybrid groupings with both a subgroup (SG4) and two standalone devices, demonstrating the algorithm's adaptive granularity to heterogeneous compute resources.

\subsection{Parallel Strategy Planner}
\label{parallel-strategy-planner}


Conventional parallelism strategies are ineffective for CEE environments (\S\ref{hierarchical-topologies}).
To address this, \sysname introduces the \textit{Planner}, which implements a \textit{compact zero-bubble pipeline architecture} that decouples weight updates from gradient computation and proactively inserts the weight updates to pipeline bubbles incurred by the slow and heterogeneous networks whenever feasible.
Building upon this foundation, \sysname further integrates a region-aware cost model alongside an automated strategy search algorithm to collaboratively generate optimized parallel execution plans.


\begin{figure}[t]
    \centering

    \subfloat[The original zero-bubble pipeline.]{\includegraphics[width=8cm]{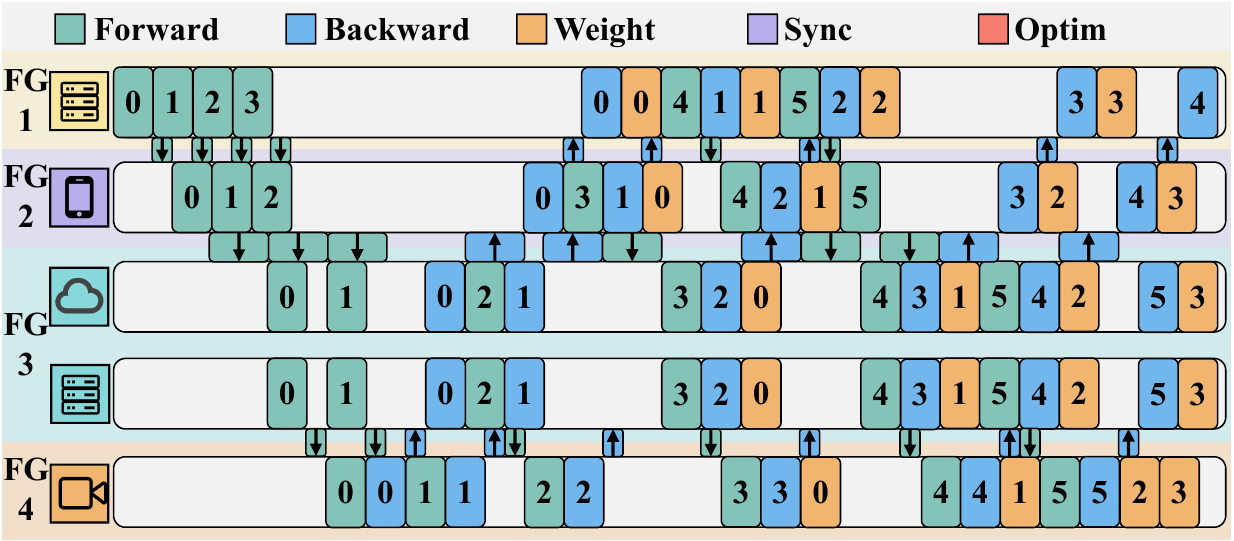}\label{fig:train-stage:origin}}
    \hfill
    \subfloat[The compact zero-bubble pipeline.]{\includegraphics[width=8cm]{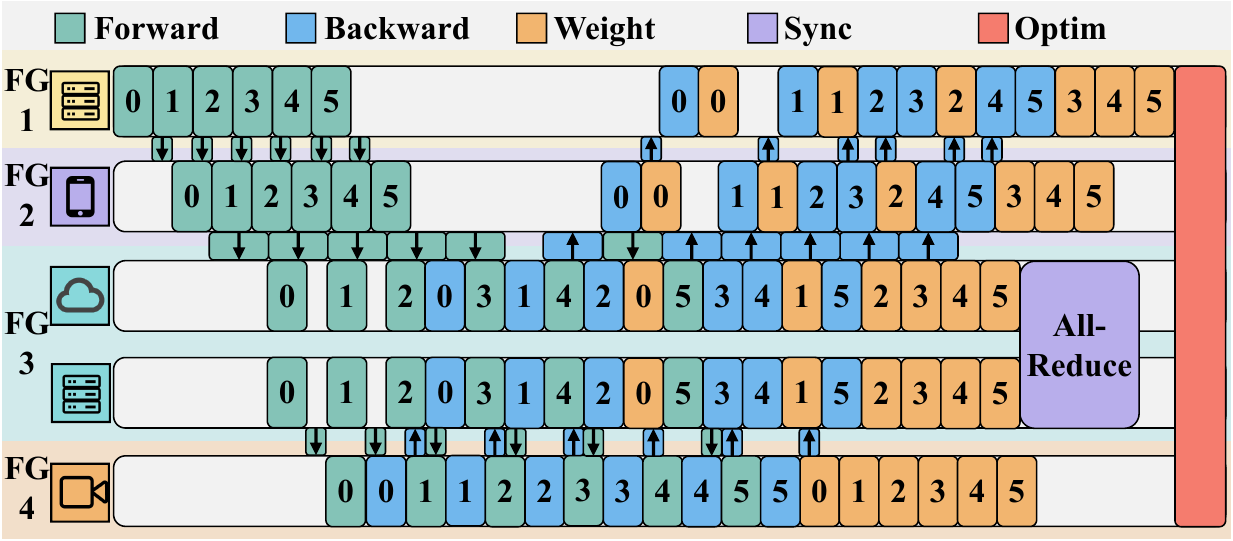}\label{fig:train-stage:greedy}}

    \vspace{-0.3cm}
    \caption{The comparison of original and compact zero-bubble pipeline. (FG: first-level network device group)}
    \vspace{-0.8cm}
    \Description{The pipeline architecture and cost modeling stage of \sysname. (G:first-level device group)}
    \label{fig:train-stage}
\end{figure}

\subsubsection{Compact Zero-bubble Pipeline Architecture}
\label{pipeline-architecture}
As that pipeline parallelism (PP) incurs the lowest communication overhead, \sysname adopts PP across first-level groups.
In particular, the \textit{Planner} implements compact zero-bubble PP, building upon the original zero-bubble pipeline design \cite{qiZeroBubblePipeline2023}, which decouples gradient computation from weight updates. This decoupling substantially enlarges the pipeline scheduling space.
However, the original zero-bubble PP assumes that inter-stage communication is faster than both forward and backward propagation computation.
This leads to a rigid execution pattern where forward passes, backward passes, and weight updates follow fixed, non-overlapping phases.
Original zero-bubble pipeline generally follows the 1F1B schedule, but it adjusts the starting points of weight update depending on the number of warm-up micro-sbatches.
However, CEE’s hierarchical cluster topology and bandwidth-constrained networks violate its fundamental assumption.
As Figure~\ref{fig:train-stage:origin} demonstrates, applying the original zero-bubble pipeline to CEE environments exposes critical limitations. 
Its fixed scheduling pattern generates excessive GPU bubbles that leave most of computing resources idle during network bandwidth-constrained periods.

As shown in Figure~\ref{fig:train-stage:greedy}, 
in compact zero-bubble pipeline, there are five operations: forward propagation, backward propagation, weight update, weight synchronization, and optimizer operations.
The Planner orchestrates operation execution through the following principle.
First, during pipeline warm-up, we greedily execute all available forward propagations, unlike the original zero-bubble pipeline that waits for backward propagation completion before proceeding.
This design exploits inter-region network conditions to maximize forward pass throughput, enabling tighter pipeline packing later.
As Figure~\ref{fig:train-stage:origin} demonstrates, while the original zero-bubble pipeline would delay forward propagation \#4 until after FG1's weight update \#0 (creating cascading delays for FG2-FG4).
Our compact zero-bubble pipeline completes all six forward passes first.
This results denser operation scheduling in downstream stages, reducing total GPU bubbles.
Second, we prioritize operations based on their dependency chains: forward propagation (highest), backward propagation (medium), and weight updates (lowest).
This hierarchy reflects backward propagation's dependence on both current and preceding stages' forward passes, while weight updates only require local gradient completion.
As Figure~\ref{fig:train-stage:greedy} demonstrates in FG2's execution: after finishing weight update \#2, the system immediately processes backward propagations \#4 and \#5 before completing remaining weight operations.
This scheduling strategy triggers FG1's backward passes earlier, effectively eliminating GPU bubbles in FG1 through proactive dependency resolution.
Finally, we employ compact weight update scheduling to exploit all residual GPU bubbles.
While prioritized operation scheduling produces tightly-packed pipelines in CEE environment, network performance difference is unavoidable.
Our solution dynamically reorganizes weight updates to fill these idle time.
As Figure\ref{fig:train-stage:greedy} demonstrates in FG3, we insert weight update \#0 between backward propagation \#2 and forward propagation \#5, then immediately resume backward propagation \#3 upon its dependency resolution, temporarily suspending non-critical weight updates \#1 and \#2.


Building atop the pipeline architecture, the \textit{Planner} partitions the model and distributes the resulting segments across first-level groups.
Furthermore, it allows heterogeneous parallelism strategies within each first-level network device group, enabling subdivision into second-level computing device groups based on hardware computing capabilities.
This flexibility allows optimal strategy selection (DP, TP, or hybrid) tailored to each second-level group's specific device characteristics and workload requirements.

\subsubsection{Cost Model}
\label{cost-model}
To find optimal parallel strategies for CEE environments, \sysname first formalizes their costs by Equation~\ref{eq:cost}.
It then tries to find a solution that minimizes the following cost ($T^*$) based on the compact zero-bubble PP.

\begin{equation}
    T^*= \min \max_{s \in \mathbb{S}} \{\sum_{i=1}^P t^f_i + B \cdot t^c_s + \sum_{i=1}^P t^l_i + AL\}
    \label{eq:cost}
\end{equation}

\noindent
$s \in \mathbb{S}$ denotes a specific pipeline stage $s$ within the all stage set $\mathbb{S}$. 
$P$ represents $s$’s predecessor stages, $t^f_i$ denotes the computation and transfer time between stage $i$ and $i+1$, and the summation represents the cost of fill stages.
$B$ denotes the micro-batch size, $t^c_s$ denotes the computing time per micro-batch in stage $s$ and their product represents the cost of running stages.
$t^l_i$ denotes the latency when predecessor stages’ computation fails to fully overlap communication (i.e., synchronization cost).
AL is the collective communication time for DP and TP within a first-level network device group.

Stage computing time ($t^c$ and $t^f$) is derived from the model’s FLOPs requirements normalized by each first-level network device group’s compute capability (obtained from the \textit{Profiler},  \S\ref{heterogeneous-devices-profiler}).
The residual latency ($t_l$) quantifies non-overlappable communication as $t^l = t_{comm} - t_{lap}$, where $t_{comm}$ is communication time between two stage and $t_{lap}$ is overlapped communication time (i.e., computing time).
Intra-group communication overhead ($AL$) is modelled as $AL = V / \min(bandwidth)$, where $V$ is communication volume and $\min(bandwidth)$ is lowest bandwidth in a first-level network device group.

The cost model effectively captures training overheads in hierarchical cluster topology, enabling accurate pipeline bubble prediction.
This formulation also provides the foundational metrics for our automated search algorithm.

\subsubsection{Search Algorithm}
\label{search-algorithm}

Leveraging the pipeline architecture (\S\ref{pipeline-architecture}) and cost model (\S\ref{cost-model}), we design an automated search algorithm that abstracts CEE's hierarchical topology, generating near-optimal parallelism strategies.




\noindent \textbf{First-level Network Device Group.}
As Equation~\ref{eq:cost} indicates, the \textit{Planner} needs to identify the pipeline stage with the highest cost among the first-level network device groups and seek to minimize it.

\begin{algorithm}[t]
    \caption{Planner first-level network device group search algorithm.}
    \label{algo:search}

    \ForEach{$b \in \mathbb{B}$}{
        \ForEach{$m \in \mathbb{M}$}{
            $\mathbb{A} \gets$ Generate the candidates with length $l$ \;
            \For{$i=0; i < \mathrm{max\_iter}; i++$}{
                $\mathbb{A} \gets$ Expand the candidates \;
                \ForEach{$a \in \mathbb{A}$}{
                    $T^*_a \gets$ Calculate the current optimization goal\;
                }

                Sort the $\mathbb{A}$ by $T^*_a$ \;
                $\mathbb{A} \gets \mathbb{A}[0:l]$

            }
        }
    }

    \Return $\mathbb{A}[0]$
\end{algorithm}

The search algorithm~\ref{algo:search} explores optimal model spliting within a four-dim space comprising batch size ($b$), microbatch size ($m$), model layers, and first-level network device groups.
It first enumerates all possible partition combinations for given $(\mathbb{B}, \mathbb{S})$ configurations (Lines 1-2), then employs beam search to efficiently navigate the subspace of model layers and first-level network device groups.
The planner initializes by randomly permuting first-level network device groups and generating $l$ candidate split set $\mathbb{A}$ through computing aware model segmentation (Line 3).
Through iterative refinement (up to max\_iter cycles), it expands the search space by (1) reordering pipeline stages for global exploration and (2) swapping adjacent layers for communication aware local optimization (Line 5). 
Each candidate split is evaluated via a cost model (Lines 6-7), with the top-l lowest-cost solutions retained for subsequent iterations (Lines 8-9), progressively converging toward the optimal split strategy.

\noindent \textbf{Second-level Computing Device Group.}
Within each first-level network device group, we partition the model according to the computing capacity of second-level computing device groups.
This is achieved through fully decoupled tensor splitting along the tensor's first and second dimensions, enabling arbitrary proportional allocation across devices to maximize computational resource utilization.
Specifically, we designed the following three partitioning strategies:

\begin{figure}
    \centering

    \subfloat[Asymmetric PP]{\includegraphics[height=4cm]{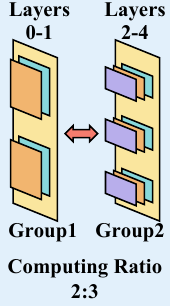}\label{fig:asymmetric:pp}}
    \hfill
    \subfloat[Asymmetric DP]{\includegraphics[height=4cm]{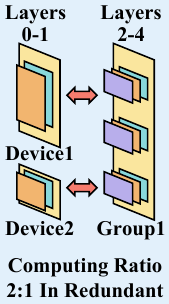}\label{fig:asymmetric:tp-and-dp}}
    \hfill
    \subfloat[Asymmetric DP and TP]{\includegraphics[height=4cm]{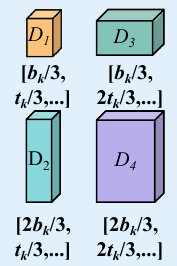}\label{fig:asymmetric:direct}}

    \vspace{-0.3cm}
    \caption{Examples of asymmetric multi-dimensional parallelism. (Group: second-level device group)}
    \vspace{-0.5cm}
    \Description{Examples of asymmetric multi-dimensional parallelism. (Group: second-level device group)}
    \label{fig:asymmetric}
\end{figure}

\textbf{1) Asymmetric PP.}
PP is also implemented among second-level device groups.
Specifically, \textit{Profiler} can obtain computing capabilities and memory size of second-level computing device group within given first-level network device group.
According to the ratio of computing capabilities between the second-level device groups, the model layers belonging to this stage are further divided into different second-level groups, forming an asymmetric PP. 
Meanwhile, the standard symmetric TP or DP are implemented within each second-level device group.
For example, as shown in Figure \ref{fig:asymmetric:pp}, the model layers 0-5 are divided into 2 second-level device groups according to their computing capacity ratio.

\textbf{2) Asymmetric DP.}
When PP partitioning encounters a performance bottleneck, such as when a second-level computing device group contains only one device with weak computing power, similar to this type of second-level computing device group, they can collaborate together to
compute the same model layer through asymmetric data splitting.
For example, as shown in Figure~\ref{fig:asymmetric:tp-and-dp}, we divide the 5-layer model into two parts: Group1 is responsible for computing 3 layers, while devices 1 and 2 (the second-level group has only one device) compute 2 layers.
Additionally, the input data is split between devices 1 and 2 in a 1:2 ratio to ensure they complete computations simultaneously, thus avoiding the PP computation bottleneck.

\textbf{3) Asymmetric TP and DP.}
As mentioned above, among devices more dimension asymmetric TP and DP are supported to achieve that, the tensor and data are divided among devices based on the alignment of the computing capabilities of individual heterogeneous devices to avoid bottleneck nodes.
For example, as shown in Figure \ref{fig:asymmetric:direct}, the computing capacity ratio of heterogeneous devices is $1:2:2:4$, therefore, for a tensor with dimensions of $[b_k, t_k]$, according to this ratio, it is divided into 4 unequal tensors, namely $[b_k/3, t_k/3]$, $[2b_k/3, t_k/3]$, $[b_k/3, 2t_k/3]$, $[2b_k/3, 2t_k/3]$, and which are assigned to corresponding devices, respectively.

\subsection{Dynamic Environment Adapter}
\label{dynamic-environment-adapter}

Network fluctuations in CEE environments introduce additional GPU bubbles throughout the training pipeline (\S\ref{network-fluctuations}).
To address this issue, the \textit{Adapter} decouples micro-sizes across pipeline stages, enabling independent size adjustment per stage to minimize waiting latency between stages.
During network fluctuations, the \textit{Adapter} dynamically adjusts the micro-batch size of affected pipeline stages to enable earlier initiation of subsequent micro-batch processing.

\begin{figure}
    \centering

    \subfloat[Adapter overview]{\includegraphics[width=4cm,height=4cm]{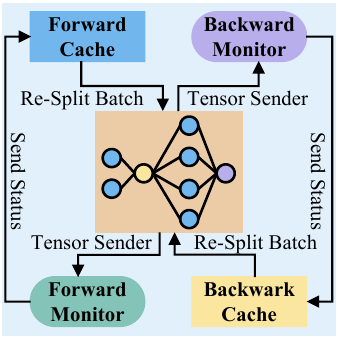}\label{fig:adapter:arch}}
    \hfill
    \subfloat[An example of Adapter]{\includegraphics[width=4cm,height=4cm]{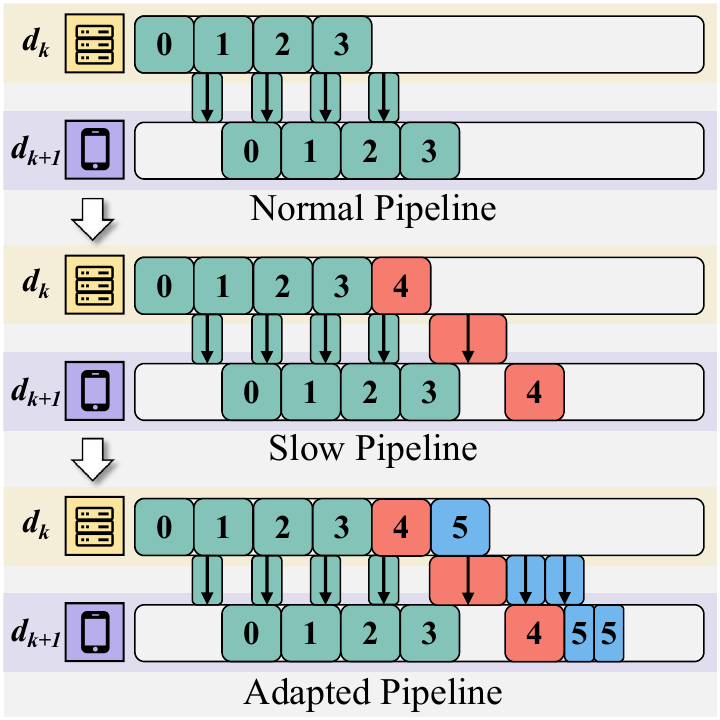}\label{fig:adapter:example}}
    
    \vspace{-0.3cm}
    \caption{The overview and example of the \textit{Dynamic Environment Adapter}.}
    \vspace{-0.3cm}
    \Description{The overview and example of dynamic environment Adapter.}
    \label{fig:adapter}
\end{figure}

The Adapter comprises two main components: the Forward/Backward Cache and the Forward/Backward Monitor (Figure~\ref{fig:adapter:arch}).
Each first-level device group includes a Forward Cache that dynamically adjusts the micro-batch size.
The Forward Monitor tracks the available network bandwidth, and based on whether the bandwidth exceeds or falls below a predefined threshold, tensors are either aggregated or re-split for forward propagation.
The Backward Cache and Monitor function in a similar manner during back propagation.
Together, these components enable dynamic adjustment of micro-batch sizes across first-level device groups.



Specifically, the Adapter adopts the following strategies to adjust the size of micro-batches dynamically.
First, to shorten the time of the filling stage, the micro-batch size of each first-level device group is reduced when the network capabilities are poor, thus completing the filling as soon as possible.

\begin{figure*}
    \centering

    \subfloat[Comparison of throughput under Setting 1.]{\includegraphics[width=8.5cm]{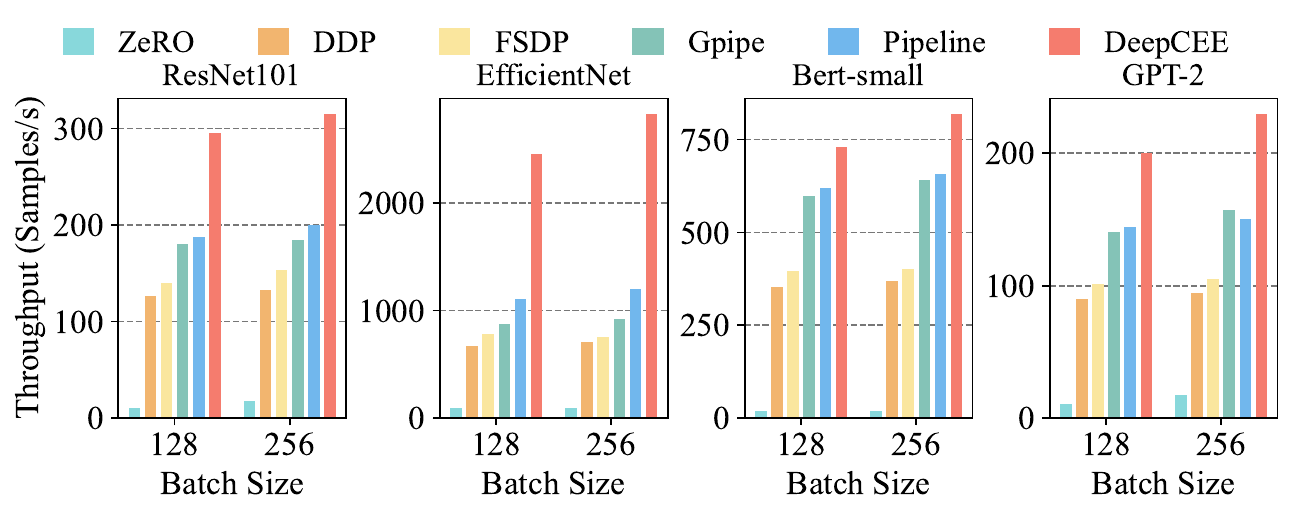}\label{fig:throghput-exp:enva}}
    \hfill
    \subfloat[Comparison of throughput under Setting 2.]{\includegraphics[width=8.5cm]{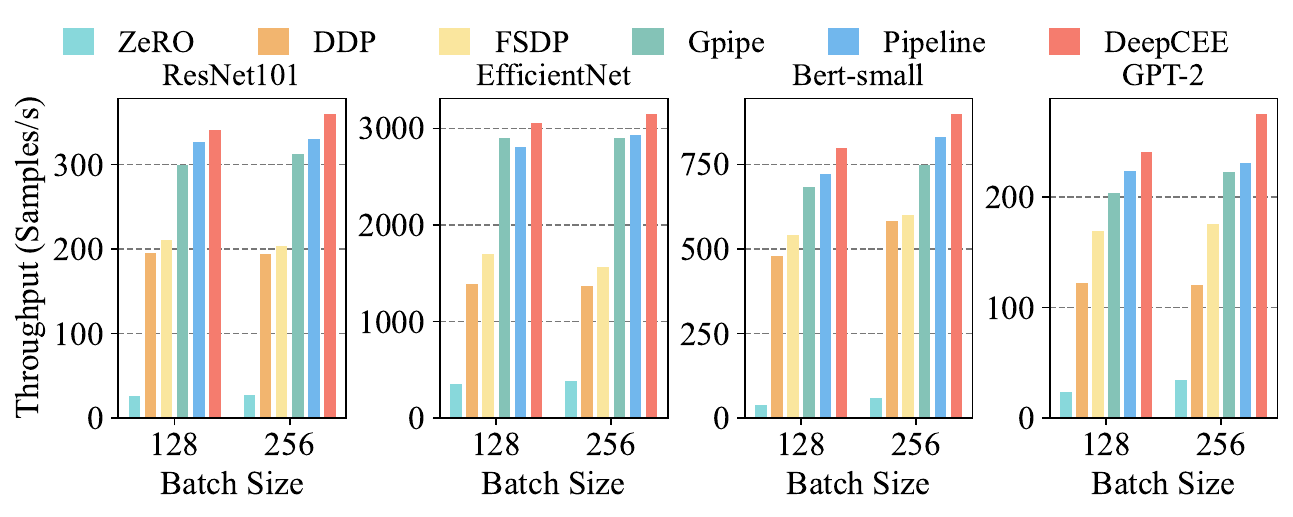}\label{fig:throghput-exp:envb}}

    \vspace{-0.3cm}
    \caption{Comparison of throughput for model training using different methods under 2 heterogeneous CEE environment.}
    \vspace{-0.6cm}
    \Description{Comparison of throughput for model training using different methods under various heterogeneous CEE.}
    \label{fig:throghput-exp}
\end{figure*}

\begin{figure*}
    \centering

    \subfloat[Comparison of resnet101.]{\includegraphics[width=4.2cm]{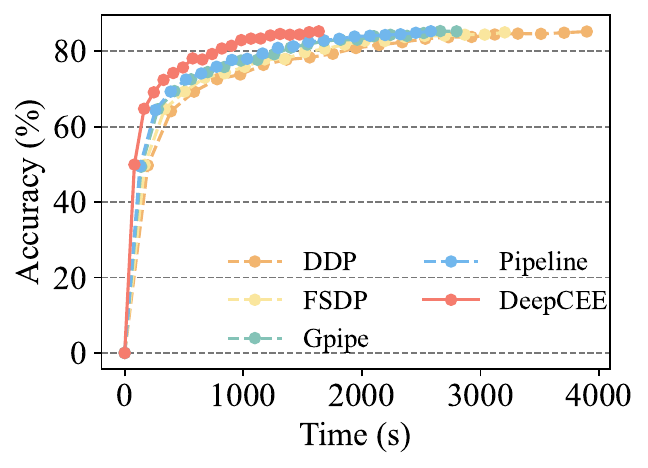}\label{fig:accuracy:resnet}}
    \hfill
    \subfloat[Comparison of EfficientNet.]{\includegraphics[width=4.2cm]{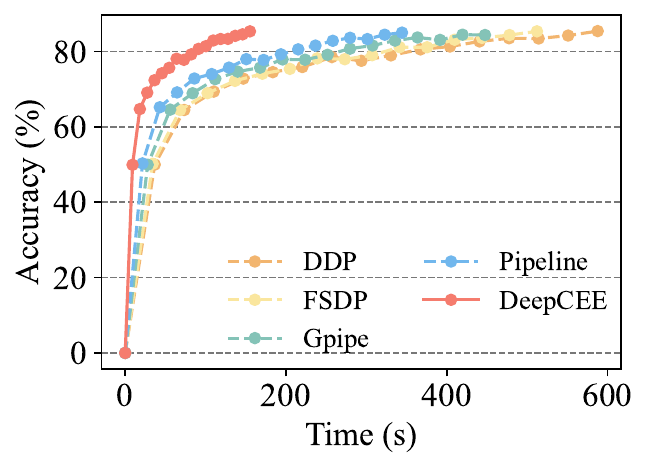}\label{fig:accuracy:eff}}
    \hfill
    \subfloat[Comparison of Bert-small.]{\includegraphics[width=4.2cm,height=3cm]{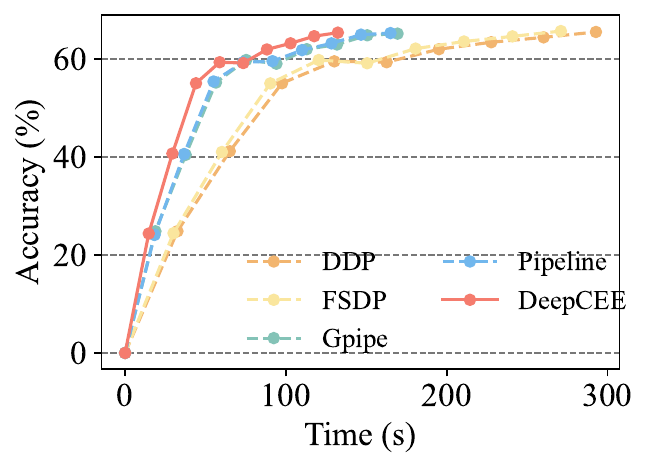}\label{fig:accuracy:bert}}
    \hfill
    \subfloat[Comparison of GPT-2.]{\includegraphics[width=4.2cm,height=3cm]{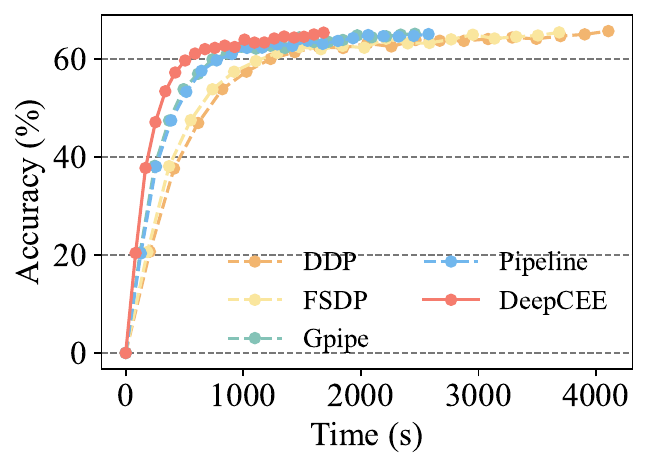}\label{fig:accuracy:gpt}}

    \vspace{-0.3cm}
    \caption{Comparison of model accuracy for model training using different methods under Setting 1.}
    \vspace{-0.3cm}
    \Description{Comparison of model accuracy for model training using different methods under environment Setting 1.}
    \label{fig:accuracy}
\end{figure*}

Second, to reduce the computing bubbles for speeding up the running stage, the Forward and Backward cache monitor the latency taken to send a micro-batch.
When they detect a significant transmission delay, the micro-batch size of the current first-level device group is reduced to minimize the waiting time for subsequent groups.
For example, as shown in Figure \ref{fig:adapter:example}, for the first-level device group $d_k$, the micro-batch size during the pipeline running process is 32.
Upon detecting a transmission delay, the size is reduced to 16, allowing the next device group $d_{k+1}$ to promptly begin its forward propagation.

Third, when the Adapter detects that the forward propagation has been completed entirely, the micro-batch size of all first-level device groups is reduced to accelerate the rest of the backward propagation.



\section{Evaluation}
\label{evaluation}

We will answer the following questions experimentally: 
What and why does \sysname improve performance compared to the state-of-the-art (SOTA) works in the CEE when training diverse models (\S \ref{end-to-end-performance}) ?
What is the effectiveness and adaptation of \sysname's parallel strategy compared to the SOTA works (\S \ref{parallel-strategy-comparison} and \S \ref{parallel-strategy-adaption})?
What is the scalability of \sysname compared to the SOTA works (\S \ref{scalability-comparison}) ?
What is the \sysname's system overhead (\S \ref{system-overhead}) ?






\subsection{Evaluation Environment}
\label{settings}

\textbf{Models and Datasets.}
We evaluate \sysname with 5 typical AI models that are widely used in the CEE, including CV models (i.e., ResNet50, EfficientNet) and NLP models (i.e., Bert-small, GPT-2).
We use Cifar-10 dataset with input $3 \times 32 \times 32$ for ResNet101, EfficientNet.
For Bert-small, GPT-2, we use the oscar-en-10k dataset with input shape $32 \times 512$. 

\textbf{Environment Setup.}
We have built a CEE experimental environment configured with 10 servers: 3 servers, each configured with 4 RTX4090 GPUs used as the cloud, 3 servers with each configurated with 2 RTX3090 GPUs used as edge servers, and 4 servers with each configured with 1 RTX3080 GPU used as end devices.
In addition, two types of network settings are set up.
For Setting 1, the network bandwidth is set to 100 Mbps for the Cloud-Edge and Cloud-End, with the remaining network bandwidth set to 1 Gbps.
In Setting 2, all network bandwidth is set to 500 Mbps.
Importantly, our method can be extended to the CEE with any resource configuration.

\textbf{Compared Methods.}
We compare \sysname with both widely-used frameworks and the SOTA methods:

\textbf{Widely-used Frameworks.}
We use PyTorch DDP \cite{liPyTorchDistributedExperiences2020}, PyTorch FSDP \cite{zhaoPyTorchFSDPExperiences2023}, DeepSpeed ZeRO2 \cite{rajbhandariZeROMemoryOptimizations2020a}, DeepSpeed Pipel-ine \cite{narayananPipeDreamGeneralizedPipeline2019}, and Gpipe \cite{huangGPipeEfficientTraining2019}as comparison systems for end-to-end performance evaluation.

\textbf{The SOTA Methods.} 
We use Alpa \cite{zhengAlpaAutomatingInter2022}, Metis \cite{umMetisFastAutomatic2024}, Hetpipe \cite{parkHetPipeEnablingLarge2020}, and Asteroid \cite{yeAsteroidResourceEfficientHybrid2024} as the SOTA comparison methods for parallel strategies evaluation.
Alpa unifies inter-operator and intra-operator parallelism through integer linear programming (ILP) and dynamic programming, achieving optimal model placement for homogeneous clusters.
Metis extends Alpa to heterogeneous cloud environments by employing depth first search to identify optimal computation partitions across heterogeneous devices.
HetPipe combines data and pipeline parallelism with communication-aware scheduling to maximize throughput on GPU/TPU clusters.
Meanwhile, Asteroid specifically targets resource-constrained edge devices, developing a pipeline evaluation method that determines optimal partitions via five dimensional search space exploration.
Our evaluation combines established open-source frameworks and custom implementations.
Specifically, we employ Alpa's native JAX implementation for its automated parallelism optimization, while constructing HetPipe's pipeline mechanism using PyTorch's distribute package.
For Metis and Asteroid, we faithfully reimplement their partition search algorithms while leveraging PyTorch to execute their generated strategies.




\subsection{End-to-End Performance Evaluation}
\label{end-to-end-performance}
This section evaluates the training throughput and time overhead of \sysname against widely-used training frameworks in achieving the specified model training accuracy.

\textbf{Training Throughput.}
As shown in Figure \ref{fig:throghput-exp}, \sysname improves training performance by 2.3-2.38$\times$ and 1.74-1.85$\times$ compared to PyTorch DDP in settings 1 and 2, respectively.
It achieves 2.05–2.11$\times$ and 1.61–1.76$\times$ performance gains compared to PyTorch FSDP in settings 1 and 2, respectively.
Similarly, \sysname delivers 1.04–1.57$\times$ and 1.13–1.71$\times$ performance improvements compared to DeepSpeed Pipeline and GPipe, respectively.
These performance gains are primarily due to \sysname’s ability to effectively address device heterogeneity and dynamic network diversity in CEE environments, which are overlooked by both PyTorch DDP, PyTorch FDSP, DeepSpeed Pipeline and Gpipe, resulting in worker imbalance and inefficient synchronization.
We further observe an intriguing phenomenon in the CEE: larger batch sizes consistently deliver higher throughput.
Specifically, our evaluations demonstrate a 6.8-16.2\% throughput improvement when increasing batch size from 128 to 256 (Figure~\ref{fig:throghput-exp}).
This stems from two key factors: (1) longer computing intervals create more opportunities for communication-computation overlap, (2) increased micro-batch counts compress pipeline warm-up phases.

\textbf{Time Overhead.}
As shown in Figure \ref{fig:accuracy}, the time overhead is evaluated with using different models when achieving the same accuracy in Setting 1.
Specifically, training terminates when ResNet-101 and EfficientNet achieve the target accuracy of 85\%, and when Bert-small and GPT-2 achieve the target BLEU of 0.65.
Compared with existing approaches, \sysname takes  55-73.5\% less time.
This is mainly thanks to \sysname can adapt to the CEE by a reasonable hybrid parallel training strategy.

However, although \sysname can get better training performance than existing methods, it does not fully translate the training throughput improvement into training acceleration, resulting in the training time increased by 6.4-8.2\% in contrast to theoretical improvement.
This occurs because our compact zero-bubble pipeline disrupts the conventional weight update sequence, forward and backward no longer strictly operate on the recent model parameters.
While such aggressive out-of-order execution improves average throughput by 68.1\% (Figure~\ref{fig:throghput-exp}), it introduces parameter instability that requires 6.4-8.2\% more training iterations to reach target accuracy (Figure~\ref{fig:accuracy}).
Nevertheless, \sysname still achieves faster end-to-end training compared to other approaches,
making this trade-off acceptable for most deployments.

\subsection{Effectiveness Evaluation of \sysname}
\label{parallel-strategy-comparison}
This section compares \sysname with SOTA frameworks, in terms of training throughput and search cost of parallel strategies. We use Setting 1 for evaluation.

\begin{figure}
    \centering

    \subfloat[Throughput comparison]{\includegraphics[width=4cm]{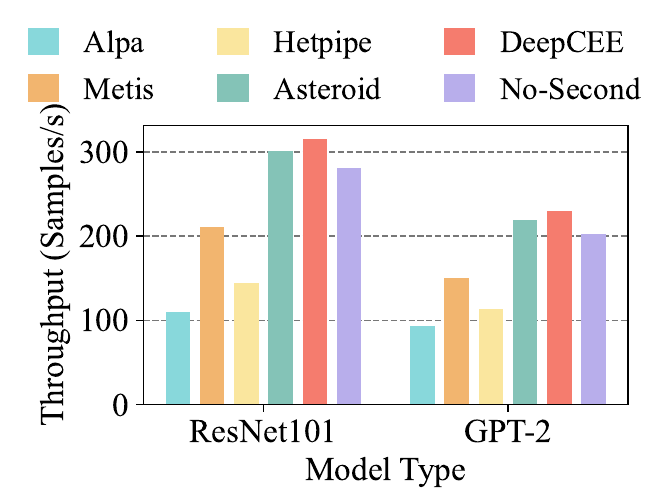}\label{fig:generate:through}}
    \hfill
    \subfloat[Strategy generation time.]{\includegraphics[width=4cm]{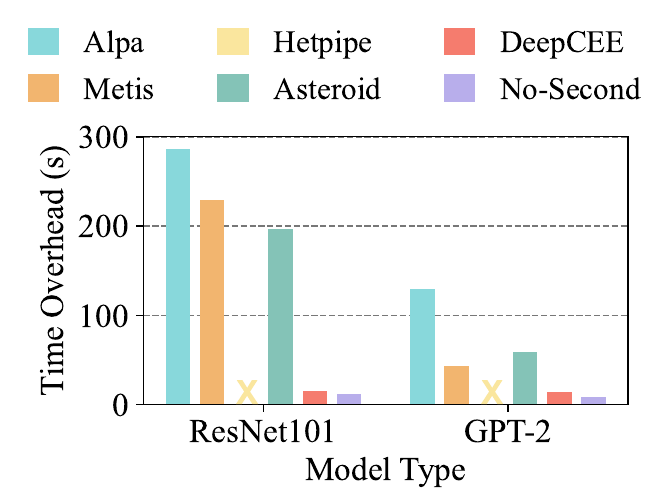}\label{fig:generate:time}}

    \vspace{-0.4cm}
    \caption{Comparison of different parallel strategy generation methods.}
    \vspace{-0.6cm}
    \Description{Comparison of different parallel strategy generation methods.}
    \label{fig:generate}
\end{figure}

\noindent \textbf{4.3.1 \ Training Throughput.}
As shown in Figure \ref{fig:generate:through}, when evaluating the training throughput of ResNet, EfficientNet, Bert-small, and GPT-2 using different auto-parallel training frameworks, \sysname can outperform existing frameworks in most cases by 1.49-2.85$\times$.

\textbf{Comparison against Alpa.}
\sysname offers 2.46–2.85$\times$ higher training throughput compared to Alpa.
While Alpa pioneered rapid parallel strategy search through unified inter-op/intra-op parallelism, combining tensor, data, and pipeline parallelism via integer linear programming.
However, its optimization targets homogeneous clusters (e.g., uniform 8-GPU nodes) and assumes static hardware conditions.
This leads to rigid, evenly partitioned model sharding that ignores device heterogeneity and network variance.
In contrast, \sysname incorporates Profiler that obtains device compute capabilities and network bandwidth, enabling dynamic load balancing adapted to real-time hardware performance.

\textbf{Comparison against Metis.}
Our experiments also compare \sysname with Metis, a variant of Alpa.
While Metis demonstrates measurable throughput improvements over Alpa in our test environment by accounting for device heterogeneity, \sysname maintains a 1.49-1.52$\times$ throughput advantage.
This performance gap stems from Metis' simplified heterogeneity model, which only addresses basic inter-cluster differences between two heterogeneous computing groups while neglecting network diversity and hierarchy.
Metis' sole optimization objective of minimizing inter-cluster communication consequently yields limited performance gains.

\textbf{Comparison against HetPipe.}
\sysname demonstrates superior training throughput, achieving 2.03-2.18$\times$ improvements over HetPipe.
HetPipe coordinates heterogeneous devices and diverse networks through communication scheduling.
But, its parameter server (PS) architecture necessitates epoch-level weight synchronization that halts training and limits throughput.
However, \sysname's fully asynchronous pipeline parallelism eliminates this bottleneck.

\textbf{Comparsion against Asteroid.}
\sysname achieves 1.04-1.05$\times$ higher training throughput than Asteroid.
Asteroid uses dynamic programming based multi-dimensional search approach to achieve similar performance.
However, \sysname's strategy more comprehensively accounts for the CEE hierarchical structures, avoiding the suboptimal parallel strategy generated by Asteroid's overcomplexed assumptions.
Furthermore, \sysname's significantly faster strategy search enables quicker training initialization, and its dynamic microbatch scheduling combined with asynchronous execution provides additional performance gains beyond Asteroid's capabilities.


\noindent \textbf{4.3.2 \ Cost Time.}
As illustrated in Figure \ref{fig:generate:time}, Alpa, Metis, and Asteroid exhibit substantially longer parallelism strategy search time compared to \sysname.
Alpa requires 286.3 seconds for ResNet101 and 130 seconds for GPT-2, primarily due to its exhaustive search across all possible device combinations, layer mappings, and hyperparameters (e.g., batch/micro-batch sizes), resulting in an $O(n! \cdot n! \cdot n^2)$ search space.
Metis reduces this overhead to 230 seconds (ResNet101) and 43 seconds (GPT-2) by pre-grouping devices into performance, balanced device groups and applying search-space pruning, though it still maintains $O(n! \cdot n^3)$ time complexity.
Asteroid further shortens the search time to 197 seconds (ResNet101) and 59 seconds (GPT-2) but remains slower than \sysname because it (1) operates on the full device set without pre-grouping and (2) incurs additional latency from runtime memory analysis and parallel strategy evaluation.

In contrast, \sysname achieves significantly faster strategy discovery through three key optimizations: (1) topology-aware device grouping that reduces the search space to near-constant complexity, (2) rapid candidate generation via predefined partitioning points, and (3) lightweight refinement using beam search.


\noindent \textbf{4.3.3 \ Profit of Second-Level Computing Device Group.}
In Figure \ref{fig:generate:through}, we evaluate the training performance when turning off \sysname's model split among second-level device groups (No-Second).
The experimental results show that the training performance of No-Second declines by 10.93\% and 12.04\% compared to \sysname when using ResNet101 and GPT-2, respectively.
This is due to the inability to fully leverage the resources of the heterogeneous computing devices when model splitting is disabled among second-level device groups.
However, its training performance is still higher than existing parallel training frameworks such as Alpa, Metis, and Hetpipe.
As shown in Figure \ref{fig:generate:time}, turning off \sysname's second-level device group search can reduce its search time of the parallel strategy by roughly 20\%. 

\subsection{Evaluating \sysname's Adaptation}
\label{parallel-strategy-adaption}

\begin{figure}
    \centering

    \subfloat[Planned geo-distributed training parallel strategy.]{\includegraphics[width=4cm]{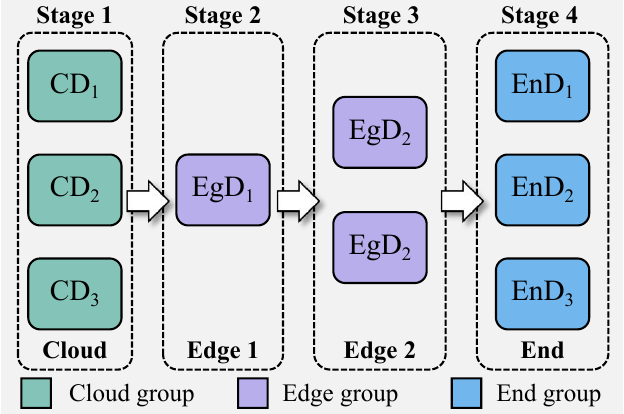}\label{fig:adapter-exp:para}}
    \hfill
    \subfloat[Comparison of the impact of network fluctuations on throughput.]{\includegraphics[width=4cm]{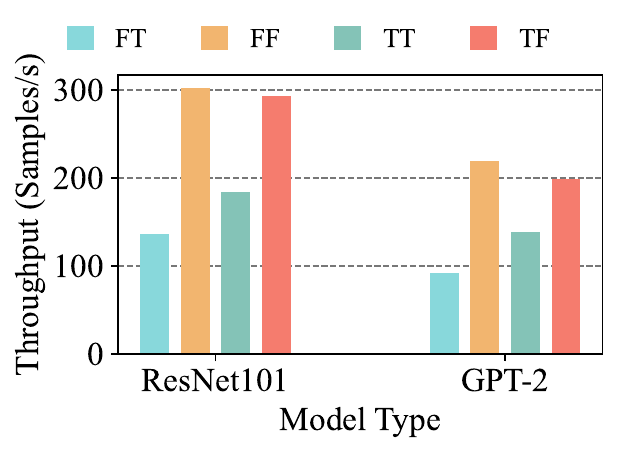}\label{fig:adapter-exp:throughput}}

    \vspace{-0.4cm}
    \caption{Evaluation results for \sysname's Adapter component.}
    \vspace{-0.6cm}
    \Description{Evaluation results for \sysname's Adapter component.}
    \label{fig:adapter-exp}
\end{figure}

As shown in Figure \ref{fig:adapter-exp}, with Setting 1, we evaluate the \sysname's adaption in the network fluctuation environment.
As shown in Figure \ref{fig:adapter-exp:para},  the ResNet101 and GPT-2 models are divided into 4 stages and spread over different 4 first-level network device groups. 
During our experiments, we add the noise to the network between 4 first-level network device groups, reducing their bandwidth by 40-60\%, to emulate the network fluctuation. 
We evaluate four scenarios formed by two binary conditions: (1) \textit{Adapter} activated or not and (2) network noise added or not.
Specifically, including \textcircled{1} with the \textit{Adapter} off and with network bandwidth degradation (FT), \textcircled{2} with the \textit{Adapter} off and without network bandwidth degradation (FF), \textcircled{3} with the \textit{Adapter} on and with network bandwidth degradation (TT), and \textcircled{4} with the \textit{Adapter} on and without network bandwidth degradation (TF).

As shown in Figure~\ref{fig:adapter-exp:throughput}, the Adaptor can provide high training throughput when facing the dynamic changing of the network. 
The FT and TT comparison in Figure~\ref{fig:adapter-exp:throughput} reveals that \textit{Adapter} improves training throughput by up to 1.35$\times$ and 1.49$\times$ for ResNet101 and GPT-2 during bandwidth fluctuations.
This acceleration from its dynamic network-aware optimization: (1) decoupling micro-batch sizes across pipeline stages, and (2) independently adjusting microbatches per stage based on real-time network latency measurements.
Under stable network status, our analysis reveals a slight performance decline when enabling Adapter.
As shown in Figure~\ref{fig:adapter-exp:throughput}, the FF and TF comparison demonstrates throughput reductions of 2.9\% (ResNet101) and 9.3\% (GPT-2), attributable to two factors: (1) smaller micro-batch sizes decreasing hardware utilization efficiency, and (2) increased protocol overhead from more frequent communication.
However, these costs can be accepted, in the whole training process including network fluctuation, Adapter can improve the throughput by  32.1-39.7\%, when offsetting its own cost mentioned above.

\subsection{Scalability Evaluation}
\label{scalability-comparison}

\begin{figure}
    \centering
    
    \subfloat[Throughput results for varying number of devices with ResNet101.]{\includegraphics[width=4cm]{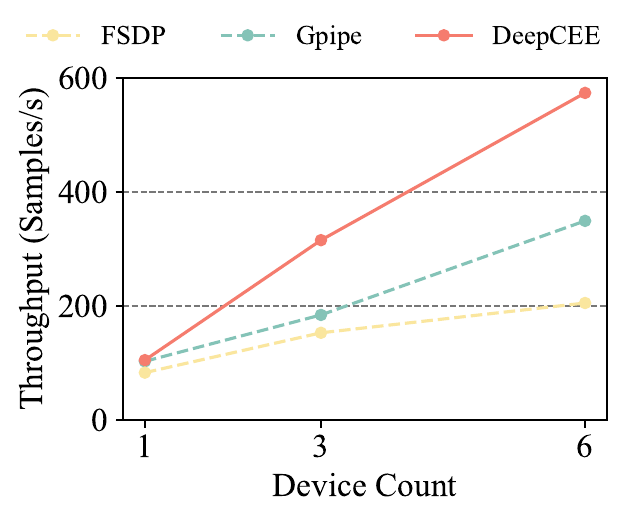}\label{fig:scalability:renet}}
    \hfill
    \subfloat[Throughput results for varying number of devices with GPT-2.]{\includegraphics[width=4cm]{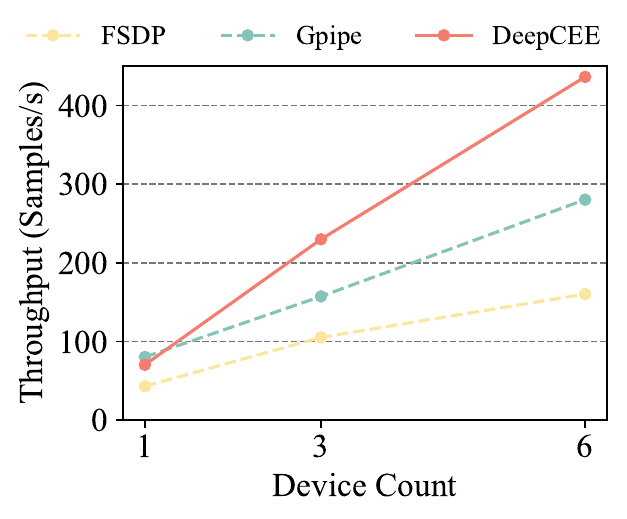}\label{fig:scalability:gpt}}

    \vspace{-0.4cm}
    \caption{Throughput results for varying numbers of devices.}
    \vspace{-0.6cm}
    \Description{Throughput results for varying numbers of devices.}
    \label{fig:scalability}
\end{figure}

As illustrated in Figure \ref{fig:scalability}, the scalability of \sysname is evaluated by examining the model training throughput under varying numbers of devices.
Based on the above configuration of the CEE, we ran this experiment by setting the server number of edge and end to be 1, 3, and 6, respectively, with the number of cloud servers being fixed, and using the network Setting 1 configuration described in \S \ref{settings}.
ResNet101 and GPT-2 models are used to be trained, and the batch size is set to be 128.

The experimental results show that \sysname's average training throughput outperforms that of PyTorch-FSDP and GPipe by 2.79$\times$ and 1.64$\times$, respectively.
That is because PyTorch FSDP requires weight synchronization across multiple nodes, where larger node counts exacerbate communication bottlenecks from stragglers.
Meanwhile, GPipe's 1F1B pipeline structure suffers from cascading inefficiencies: as model stages increase, (1) slow networks affect more stages, and (2) GPU bubble time grows proportionally with device count, significantly limiting throughput growth.
In contrast, \sysname's compact zero-bubble pipeline achieves superior scaling by: (1) decoupling weight updates to scheduling, and (2) bubble-padded execution that maximizes device resource utilization.
Our approach reduces GPU idle time and demonstrates better performance scaling with increasing devices compared to existing methods.

\subsection{System Overhead}
\label{system-overhead}

Next, we evaluate \sysname's overhead imposed by the Profiler, Planner, and Adapter components, respectively.

\textbf{The Overhead of the Profiler.}
Profiler's overhead is comprised of two parts.
First, there is the overhead originated from network capability evaluation. We employ CCL communication to evaluate network capacities, and in our current experimental setup, one cycle of network evaluation takes 15-33 seconds.
When 10 rounds of network evaluation are executed, which takes around 143 seconds.
Second, there is another overhead from computing capability evaluation. We use a medium-scale neural network to evaluate computing capability, and fitting functions are utilized to acquire the computing capability of other scale models. In the current configuration, computing capability evaluation takes about 83-159 seconds.

\textbf{The Overhead of the Planner.}
The main overhead in Planner is the search overhead for finding the optimal parallel strategy, which has been described in full in \S \ref{parallel-strategy-comparison}.
Specifically, Planner uses a dictionary to cache the results of past searches to accelerate the search process, resulting in an additional storage complexity of $O(n!)$, where $n$ is the number of first-level device groups. 
The memory overhead of the Planner is marginal, with less than 1~MB through our experimental results.

\textbf{The Overhead of the Adapter.}
The primary overhead of Adapter is caused by monitoring network transmission performance and identifying network fluctuations.
The Adapter employs timers to monitor network performance, and each timer brings tiny CPU calculation and memory overhead.
To detect network bandwidth changes, the time spent on each 20 micro-batch transmission is recorded, so the algorithm with time complexity $O(n)$ is used to determine whether there are any changes, where $n$ is the number of first-level device groups.

Overall, these are shown to be that the time and storage overhead imposed by \sysname is lightweight and causes minimal impact in model training.

\section{Related Work}
\label{related-work}

There has been extensive research on distributed training in both industry and academia.

\textbf{Single Datacenter Distributed Trainging.}
Model training within a single datacenter has been extensively studied, with parallel strategy \cite{liPyTorchDistributedExperiences2020, zhaoPyTorchFSDPExperiences2023, rajbhandariZeROMemoryOptimizations2020a, narayananPipeDreamGeneralizedPipeline2019, huangGPipeEfficientTraining2019, mohan2021analyzing, zhang2023MixPipeEfficientBidirectional, zhai2023SmartMoEEfficientlyTraining}, hardware-software co-design \cite{shen2024ASHLAdaptiveMultiStage, zhang2023AcceleratingGNNTraininga, xuSoCFlowEfficientScalable2024, mo2024HeetAcceleratingElastic, maiKungFuMakingTraining}, and fault tolerance for large scale model training \cite{mohan2021checkfreq, stratiPCcheckPersistentConcurrent2025} being the key research focuses.
For example, SmartMoE\cite{zhai2023SmartMoEEfficientlyTraining} achieves efficient MoE parallelism and training, KungFu \cite{maiKungFuMakingTraining} enables adaptive training in cloud datacenters, and CheckFreq \cite{mohan2021checkfreq} implements fine-grained checkpointing.

\textbf{Geo-distributed Training.}
MoDNN \cite{maoMoDNNLocalDistributed2017} and CoEdge \cite{zengCoEdgeCooperativeDNN2021} divide DNN workloads dynamically based on device computing capabilities, allowing for collaboration with edges.
DeepThings \cite{zhaoDeepThingsDistributedAdaptive2018} uses FTP and distributed work theft techniques to enable collaboration between IoT devices and edge clusters.
EDDL \cite{haoEDDLDistributedDeep2021} and AutoSF \cite{yangAutoSFAdaptiveDistributed2024} adjust to changing edge resources and optimize aggregation structure and frequency.
FTPipeHD \cite{chenFTPipeHDFaultTolerantPipelineParallel2024} dynamically optimizes partition locations and presents a weight redistribution mechanism to address the fault issue during training.

\textbf{Auto Parallel Strategy Generation.}
Several works, like AMP \cite{liAmpAutomaticallyFinding2022} and Alpa \cite{zhengAlpaAutomatingInter2022}, use automated tensor or operator-level parallelism on homogeneous GPUs.
Merak \cite{laiMerakEfficientDistributed2023} and Galvatron \cite{miaoGalvatronEfficientTransformer2022} can accomplish 3D parallel automation for big fundamental models on homogenous GPUs.
Metis \cite{umMetisFastAutomatic2024} automates the design of heterogeneous parallel methods and optimizes them on heterogeneous GPUs with a large parallel strategy search space.
Asteroid \cite{yeAsteroidResourceEfficientHybrid2024} automatically develops HPP strategies for multi-edge models, allowing for automated training of heterogeneous edge hardware.

\section{Conclusion}
\label{conclusion}

This paper proposes \sysname, a geo-distributed model training system, which is a communication-centric training framework to tackle the issue of slow and unstable inter-region networks in the cross-region training. Specifically, \sysname can provide a compact, zero-bubble pipeline parallelism, with automatically deriving optimal parallel strategies, and react to network fluctuations in real time. The experimental results show that \sysname can provide an efficient training performance in contrast to the SOTA methods and existing parallel training frameworks in the CEE environment.



\bibliographystyle{ACM-Reference-Format}
\bibliography{main}

%









\end{document}